\documentclass[universe,article,accept,pdftex,moreauthors]{Definitions/mdpi} 

\newcommand{\rthis}[1]{\textcolor{black}{#1}}
\DeclareUnicodeCharacter{2061}{}

\firstpage{1} 
\pubvolume{1}
\issuenum{1}
\articlenumber{0}
\pubyear{2025}
\copyrightyear{2024}
\datereceived{ } 
\daterevised{ } 
\dateaccepted{ } 
\datepublished{ } 
\hreflink{https://doi.org/} 

\Title{ A test of Amati relation using HII galaxy distances}

\TitleCitation{Title}

\Author{Rikiya Okazaki$^{1}$ and Shantanu Desai$^{2,*}$}

\AuthorNames{Rikiya Okazaki, Shantanu Desai}

\address{%
$^{1}$ \quad Department of Mathematical Engineering and Information Physics, 
            The University of Tokyo, Tokyo, Japan; r.okazaki663@gmail.com\\
$^{2}$ \quad Department of Physics, IIT Hyderabad, Kandi, Telangana-502284, India; shntn05@gmail.com (S.D.)}

\corres{Correspondence:  shntn05@gmail.com}

\abstract{We use model-independent luminosity  distances of 186 HII galaxy observations to address  the circularity problem in the Amati relation for Gamma-ray Bursts (GRBs). For this purpose, we used Artificial Neural Network based interpolation  to reconstruct the luminosity distance corresponding to the GRB redshift.
We then use two independent  GRB datasets to test the robustness of the  Amati relation at redshifts below $z=2.6$.  Our best-fit  Amati relation parameters   are consistent for the same datasets to within $1\sigma$. The intrinsic scatters which we  obtain for the two datasets of about 28\% and 35\%,  are  comparatively larger. This implies that the Amati relation using HII galaxies as distance anchors cannot be used as a probe of precision cosmology. }


\begin{document}




\section{Introduction}
Gamma-ray bursts (GRBs) are short-duration single-shot events, which have been detected over a broad energy range from  keV to TeV~\cite{Kumar}.
They are generally divided into two categories: short and long, based on whether  their duration is less than or greater than two seconds~\cite{Kouv93}. Long-duration GRBs are linked to  core-collapse supernova~\cite{Bloom}, while  short GRBs originate from binary neutron star mergers~\cite{Nakar}.  However, numerous exceptions to this general rule exist~\cite{Anastasia} (and references therein).

Over the past twenty years, GRBs have been suggested  as potential standard candles using observed correlations among various  GRB observables~\citep{Ito,Delvecchio,DainottiAmati,Moresco,Luongo21,Dainotti22,Bargiacchi}. The most widely  studied correlation is the Amati relation, which posits a correlation between the spectral
peak energy in the GRB rest frame and the equivalent isotropic radiated energy~\cite{Amati02,Amati06}. 
However, because of the scarcity  of GRBs at low redshifts, the Amati relation can only be probed  after assuming a cosmological model~\cite{Collazzi,Moresco,Ito}.

To avoid the aforementioned circularity problem, two methodologies have been used in the literature. The first approach  is to simultaneously constrain both the  cosmological model parameters and  the Amati correlation parameters~\cite{Amati08,Khadka20,Khadka21,CaoRatra24,CaoRatra25}. Alternately, 
several supplementary  probes have   been employed to obtain model-independent estimates of distances associated with  GRB redshifts,  such as Type 1a SN~\cite{LiangAmati,Liang22,Kodama08,Demianski17,Liu22copula,Huang25}, Cosmic chronometers~\cite{Montiel,Amati19,LuongoOHD,Luongo,Jain23,Wangchronometer,Huangchrono},  Baryon Acoustic Oscillation $H(z)$ measurements~\cite{Luongo}, galaxy clusters~\cite{Gowri},  X-ray and UV luminosities of quasars~\cite{Dai21}. In a similar vein, we use the HII galaxy distances to study the efficacy of the  Amati relation, without relying on an underlying cosmological model.

HII galaxies  are compact galaxies undergoing intense episodes of star formation.
The H$\beta$ luminosity  for these galaxies has been found to be  strongly correlated with the ionized gas velocity dispersion ($\sigma(H \beta)$) with  negligible  scatter~\cite{Terlevich81}.  
Therefore, observations of the  H$\beta$ flux ($F(H \beta)$)  for 
these galaxies can be used as model-independent probes of cosmic expansion~\citep{Siegel05,Chavez12,Mania12}. Most recently, they have been used to compare the $R_h=ct$ cosmological model with  the $\Lambda$CDM~\cite{Melia25} and a test of cosmic distance duality relation~\cite{Zheng25}. 
The luminosity distance ($D_L$)  and distance modulus ($\mu$) can be obtained from HII galaxy observables  ($\sigma(H \beta)$  and $F(H \beta)$)  as follows~\cite{Zheng25} (and references therein):
\begin{equation} 
    \mu = 5 \log_{10} \left(\frac{D_L}{\mathrm{Mpc}}\right) + 25
    = 2.5\left[\alpha \log_{10} \left(\frac{\sigma(H \beta)}{\mathrm{km}~\mathrm{s}^{-1}}\right)
    - \log_{10} \left(\frac{F(H \beta)}{\mathrm{erg}~\mathrm{s}^{-1}~\mathrm{cm}^{-2}}\right) + \beta\right] - 100.2,
    \label{eq:mu}
\end{equation}
where $\alpha$ and $\beta$ are empirical constants characterizing the slope and intercept of the regression relation between H$\beta$ luminosity and $\sigma(H\beta)$. The  constants ($\alpha$ and $\beta$) have been shown to be agnostic to the underlying cosmological model. We use the values
 $\alpha = 33.72 \pm 0.12$ and $\beta = 4.64 \pm\
 0.10$~\cite{Melia25}. \rthis{However, we should point out that these values of $\alpha$ and $\beta$ were  obtained in ~\cite{Melia25} from a combined fit involving both the cosmological parameters  and    $\alpha$,$\beta$, using a total of 231 sources up to a maximum redshift of $z=7.43$. Our analysis, in contrast, is primarily  restricted to redshifts up to  $z=2.6$. Since the number of additional datapoints beyond $z$ of 2.6 is still relatively small, compared to the sample analyzed, using the best-fit values from  ~\cite{Melia25} should not significantly affect our results. Nevertheless, we shall defer a joint fitting for the cosmology, $L-\sigma$ and Amati relations for  the combined (GRB + HII galaxies) dataset to a future work.}
 The luminosity distance to the  GRBs is then obtained using the above distances to HII galaxies  as anchors. 

The organization of this manuscript as outlined below.  We describe the GRB and HII galaxy data used  for our analysis in Sect.~\ref{sec:data}. Our results are discussed in Sect.~\ref{sec:results}. We conclude in Sect.~\ref{sec:conclusions}.

\section{Datasets}
\label{sec:data}

\subsection{GRB data}
The GRB datasets used in our analysis are the same as those used in ~\cite{Gowri} (G22, hereafter) and also~\cite{Dai21}. 
 The first dataset (referred to as  A220 dataset~\cite{Khadka21}),  consists of  220 long GRBs   spanning the redshift range $0.0331 \leq z  \leq 8.20$~\cite{Khadka21}. For this dataset, the redshift, peak energy of thr GRB in its rest frame ($E_p$), and GRB bolometric fluence ($S_{bol}$) have been provided. The second data set (referred to as the D17 data set)~\cite{Demianski17,Demianski2} consists of 162 long GRBs in the redshift range $0.125 \leq z \leq 9.3$. For this data set, GRB redshift, distance modulus (using Type Ia Supernovae as anchors), $E_p$, and $E_{iso}$ have been provided. For both these datasets, $1\sigma$  uncertainties are provided for each of the observables. 

\subsection{HII galaxy data}
The HII galaxy data set we  considered for this analysis  consists of a sample of 195 
sources. This includes a sample of 181 HII galaxies (HIIGx) in the redshift range $0.01 <z< 2.6$~\citep{gonzalez} and 14 newly discovered HIIGx samples~\citep{Llerena23,DeGraaf}, including 5 from JWST. The redshift range of the entire sample spans $0.0088 \leq z \leq 7.43$. However, there are very few sources located beyond redshift of 2.6. Therefore, we restrict our analysis upto a maximum redshift $z=2.6$, which yields a total of 186 HII galaxy distances.

\section{Analysis and Results}
\label{sec:results}
\begin{figure}
\centering
\includegraphics[width=0.6\textwidth]{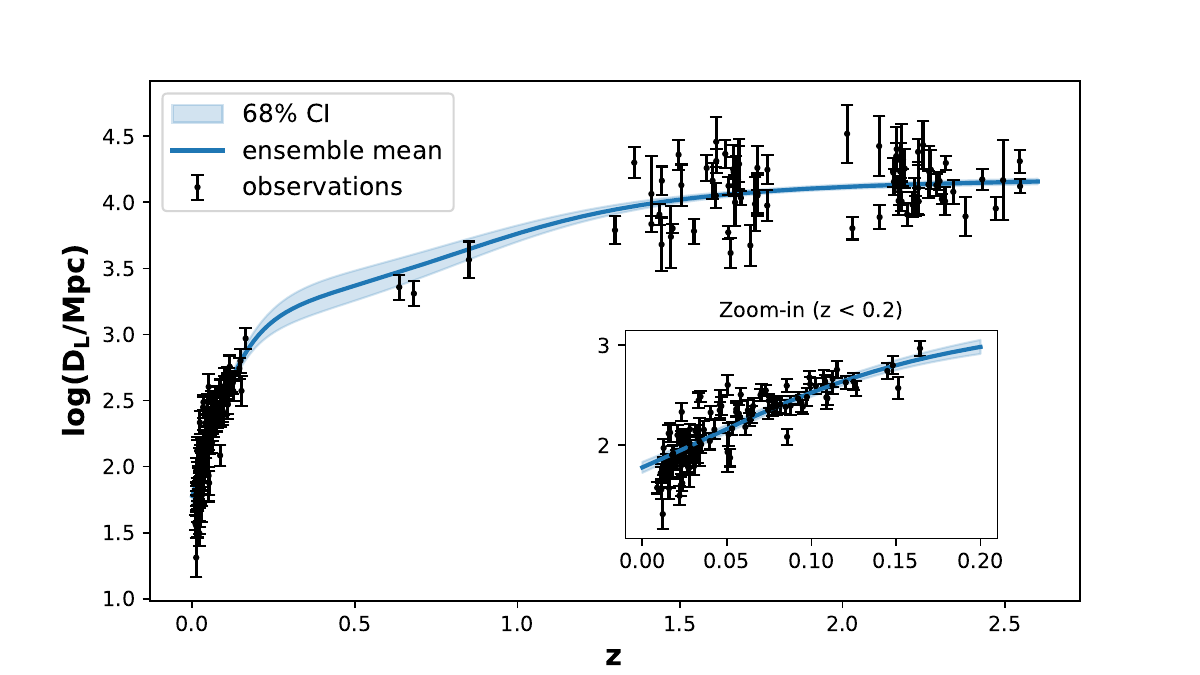}
\caption{Non-Parametric reconstruction of $D_L$  using ANN-based regression from HII Galaxy measurements upto $z$ of 2.6. 
The blue shaded region shows the 68\% allowed region. }
\label{fig1}
\end{figure}

The Amati relation can be written as a  linear regression relation between the logarithm of isotropic equivalent energy ($E_{iso}$) and GRB peak energy in the rest frame ($E_p$):
 \begin{equation}
 y = a x  + b
 \label{eq:amati}
 \end{equation}
 where $y \equiv \log (E_p (keV))$ and 
 $x \equiv \log (E_{iso} (erg))$.
 It should be noted  that $E_p$ is connected to the GRB peak energy in the observer frame through the relation $E_p= E_p^{obs}(1+z)$. $E_{iso}$ is related to the bolometric fluence ($S_{bolo}$)  according to: 
 \begin{equation}
 E_{iso}=4 \pi d_L^2 S_{bol} (1+z)^{-1},
 \label{eq:eiso}
 \end{equation}
We now obtain a model-independent estimate of $D_L$ from HII galaxy observables using Eq.~\ref{eq:mu}. 
We then need to reconstruct   $D_L$  at any intermediate redshifts. For this purpose, we use Artificial Neural Networks (ANN),  a non-parametric regression technique widely used in Astrophysics~\cite{Ball10}. A summary of other interpolation methods used for calibrating the Amati relation and comparisons with ANN can be found in ~\cite{Huang25}.
The ANN is composed of  an input layer,
several hidden layers, and  an  output layer. During each
layer, a linear transformation is applied to the input vector
from the previous layer followed by a non-linear activation function a simple mathematical operation that enables the network to learn complex, non-linear relationships), which then gets propagated to the next layer. 
For our analysis, we use the {\tt Sigmoid} activation function for all the hidden layers followed by  the identity function for the output layer. The network is implemented using {\tt pytorch}.
To perform the regression while robustly accounting for both the uncertainty arising from our finite sample size and the uncertainty propagated from  individual measurement errors, we employed ensemble learning based on bootstrapping and data perturbation~\cite{Efron1993bootstrap}, as explained below. 
We trained $K$ independent ANNs. 
The training set for each $k$-th network, $\mathcal{S}'_k$, was generated as follows:
\begin{enumerate}
    \item A bootstrap sample $\mathcal{S}_k$ was drawn with replacement from the original dataset $$\mathcal{S} = \left\{\left(z^i,\ \log D_L^i,\ \sigma_{\log D_L}^i \right)\right\}_i$$, where $\sigma_{\log D_L}^i$ denotes the standard deviation of $\log D_L^i$
    \item A perturbed dataset $\mathcal{S}'_k$ was created by sampling new target values $(\log D_L^j)'$ for each point $\left(z^j, \log D_L^j, \sigma_{\log D_L}^j\right)$ in $\mathcal{S}_k$ from a Gaussian distribution reflecting its measurement error: $$(\log D_L^j)' \sim \mathcal{N}(\log D_L^j, (\sigma_{\log D_L}^j)^2)$$
\end{enumerate}
Each ANN was then trained on its unique dataset $\mathcal{S}'_k$. For any new input $z$, the final predicted value $\log D_L(z)$ is the mean of the $K$ network outputs, and the total propagated uncertainty $\sigma_{\log D_L}(z)$ is its standard deviation.
The hyperparameters (structural settings of the model determined prior to training, such as the learning rate or the number of hidden neurons) were determined through cross-validation before the learning.
The ANN therefore enables us to reconstruct $D_L$ at any redshift $z$. This reconstructed value of $D_L$ upto redshift of 2.6 can be found in Fig.~\ref{fig1}. \rthis{We also note that there is a redshift gap between $z$ of 0.164 and 0.634. Furthermore, Ref.~\cite{CaoRatra23}  found evidence for an evolution in the $L-\sigma$ relation  between the low-redshift sample (107 data points, $z=0.0088-0.164$) and the high-redshift sample (74 data points, $z=0.634-2.545$. }
Beyond redshift of 2.6, we have very few HII galaxies, making it difficult to do the interpolation and hence we ignore these galaxies.

The ANN-based interpolation described above allows us to  reconstruct $D_L$ for both the A220 and D17 GRB dataset. We then re-estimate $E_{iso}$ using Eq.~\ref{eq:eiso} and then carry out the analysis for the Amati relation.
To calculate the best-fit parameters of the Amati relation, we use Bayesian regression and  maximize the same likelihood as that in G22:
\begin{eqnarray}
-2\ln L &=& \large{\sum_i} \ln 2\pi\sigma_i^2 + \large{\sum_i} \frac{[y_i-(ax_i+b)]^2}{\sigma_i^2 (a^2+1)} .
\label{eq:eq8}  \\
\sigma_i^2 &=& \frac{\sigma_{y_i}^2+a^2\sigma_{x_i}^2}{a^2+1}+\sigma_{s}^2 .
\end{eqnarray}
Here, $\sigma_x$ and $\sigma_y$ represent the uncertainties in $\log (E_{iso})$ and $\log (E_p)$, respectively, which are  obtained using error propagation; and $\sigma_{s}$ denotes the  intrinsic scatter, which characterizes the tightness of the relation. We used the same priors as those in G22, consisting of  uniform priors on $a$ and $b$, and log-uniform priors on $\sigma_{s}$: $a \in [0,1]$, $b \in [-30,-10]$, $\sigma_s \in [10^{-5}, 1]$. We sample the posterior using the {\tt emcee} MCMC sampler~\cite{emcee}. 

\begin{figure*}
\centering
\includegraphics[width=\textwidth]{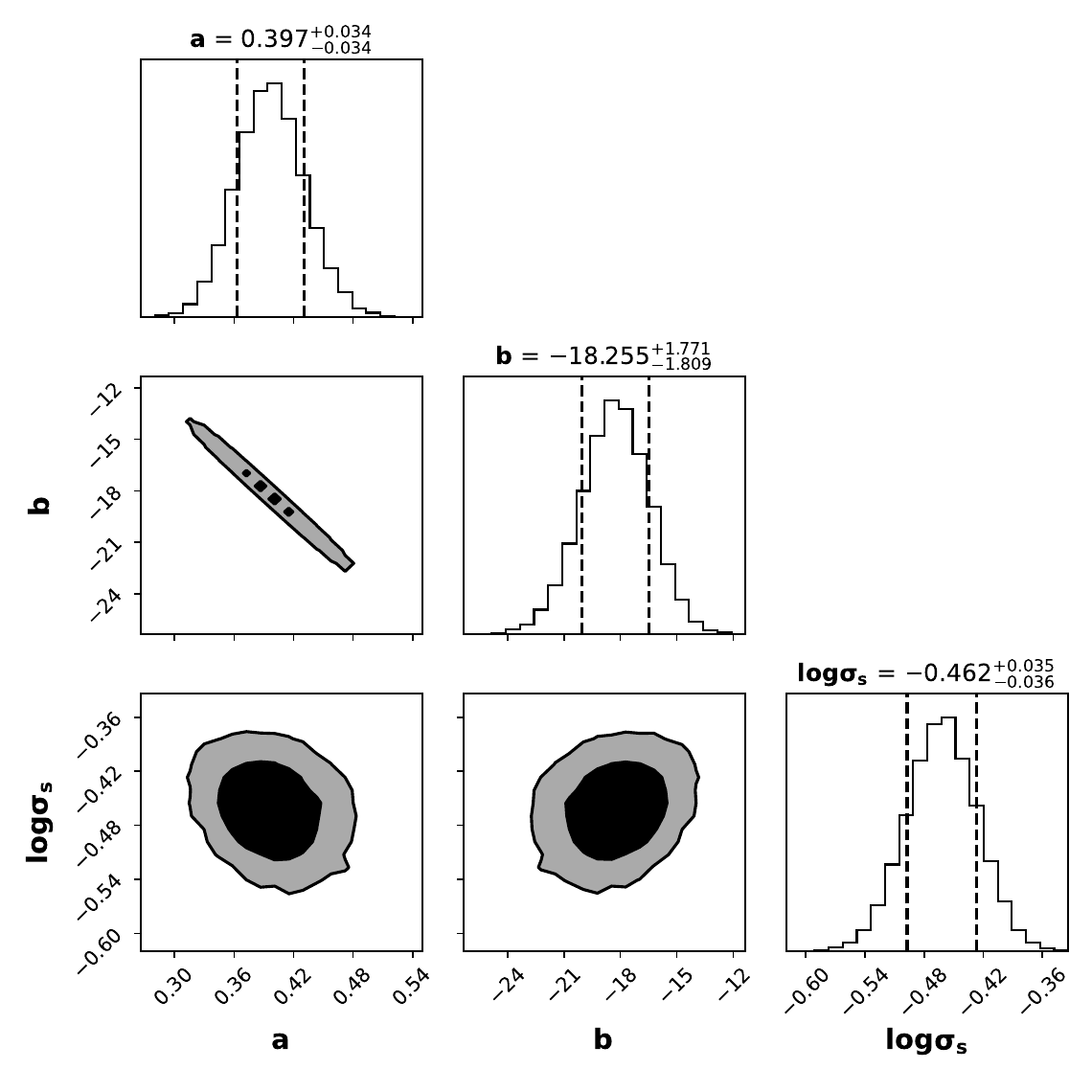}
\caption{Marginalized 68\%  and 95\% credible intervals for  $a$, $b$, and $\ln \sigma_s$ (cf. Eq.~\ref{eq:amati}) for the  subset of D17  GRBs having $z<2.6$. The dashed vertical line in the marginalized plots for each parameter  shows the 68\% uncertainty. }
\label{fig2}
\end{figure*}

\begin{figure*}
\centering
\includegraphics[width=\textwidth]{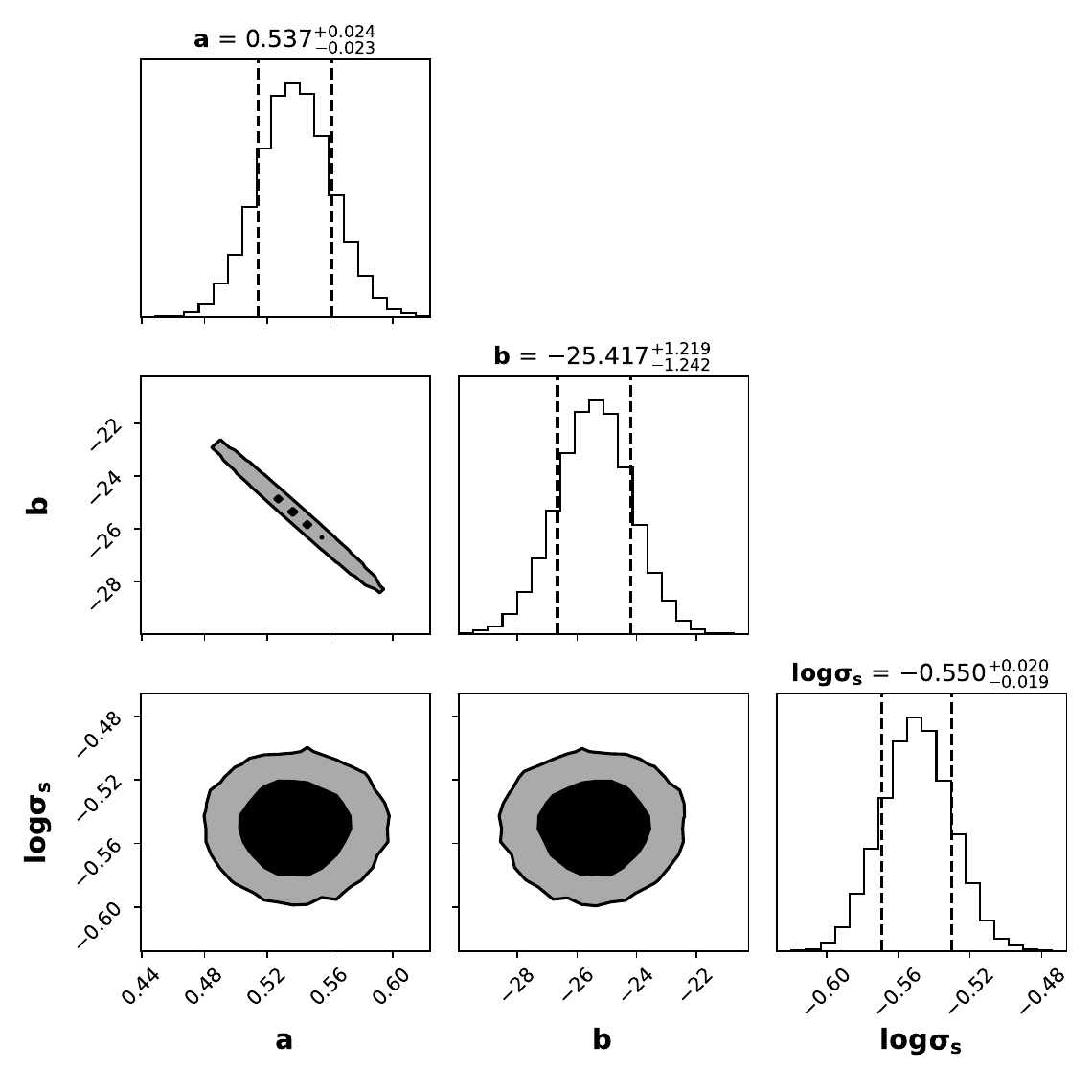}
\caption{Marginalized 68\%  and 95\% credible intervals for  $a$, $b$, and $\ln \sigma_s$ (cf. Eq.~\ref{eq:amati}) for the  subset of A220 GRBs having $z<2.6$. The dashed vertical line in the marginalized plots for each parameter  shows the 68\% uncertainty. }
\label{fig3}
\end{figure*}

\begin{figure}
\centering
\includegraphics[width=0.5\textwidth]{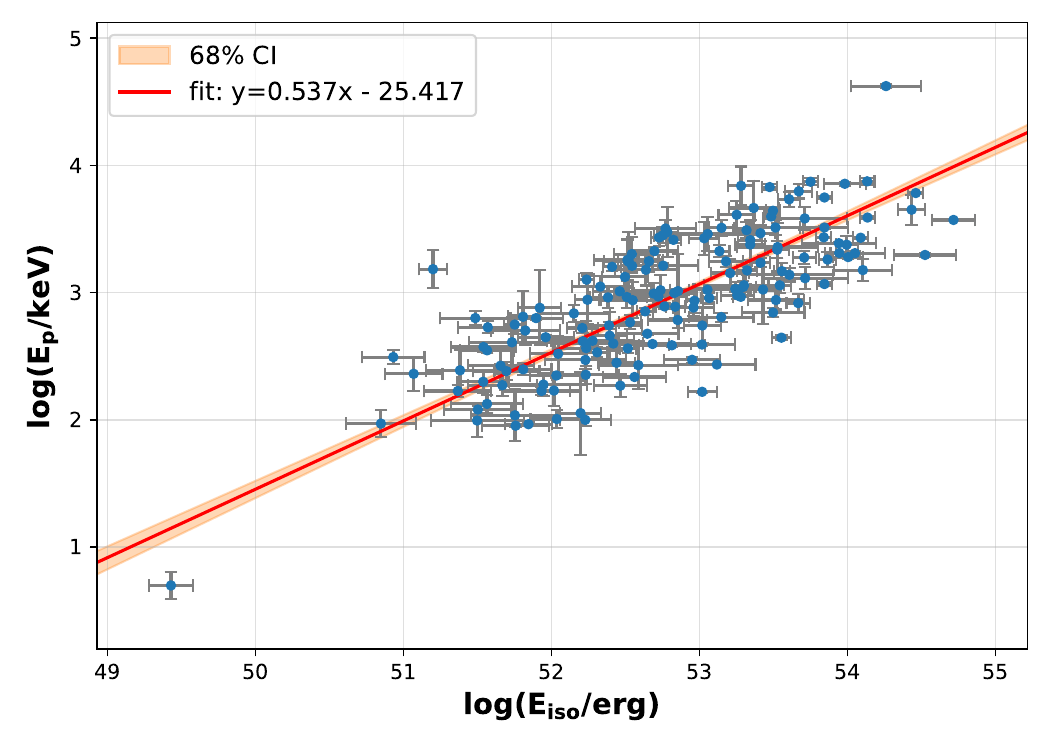}
\caption{The best-fit regression line (red line) and the $1\sigma$ confidence interval (shaded region), derived from the MCMC posterior samples along with the data for A220 dataset having $z<2.6$. The mean and standard deviation are computed at each $x$-coordinate from the ensemble of all posterior regression lines. }
\label{fig4}
\end{figure}

\begin{figure}
\centering
\includegraphics[width=0.5\textwidth]{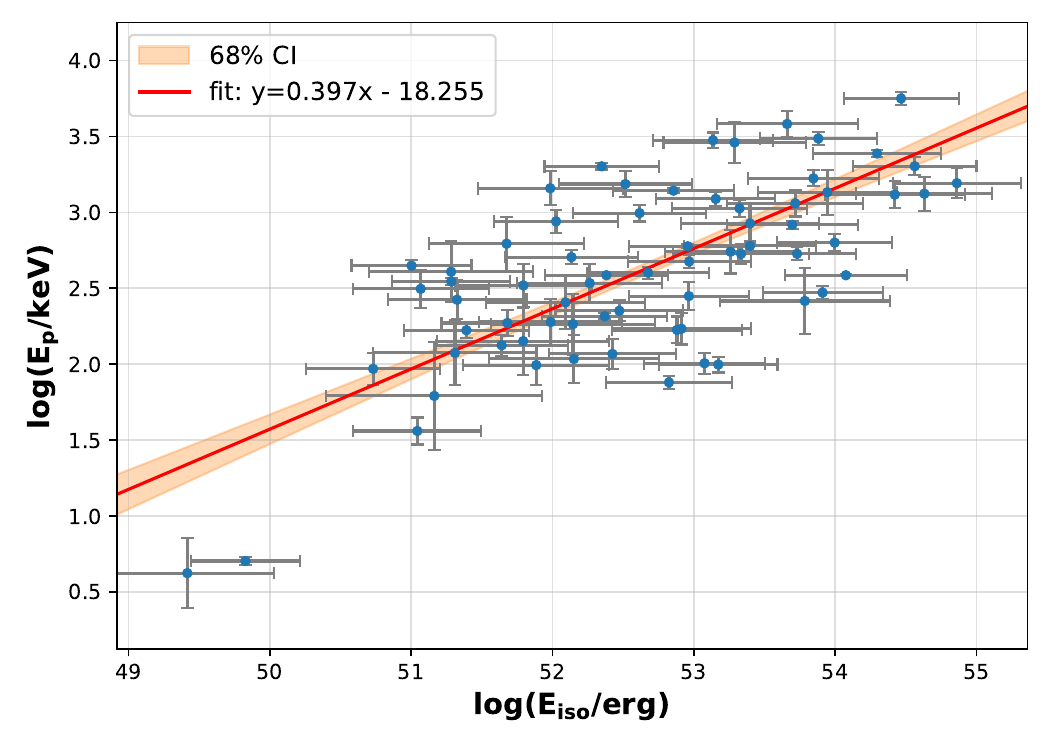}
\caption{The best-fit regression line (red line) and the $1\sigma$ confidence interval (shaded region), derived from the MCMC posterior samples along with the data for D17 dataset having $z<2.6$. The mean and standard deviation are computed at each $x$-coordinate from the ensemble of all posterior regression lines.}
\label{fig5}
\end{figure}

The marginalized 68\% and 95\% credible interval plots for the Amati relation parameters along with the intrinsic scatter can be found in Fig.~\ref{fig2} and Fig.~\ref{fig3} for the D17 and A220 GRB datasets, respectively. The scatter plots for $E_p$ versus $E_{iso}$ along with the best-fit Amati relation for  A220 and D17 dataset can be found in Fig.~\ref{fig4} and Fig.~\ref{fig5}, respectively.  A tabular summary of our results can be found in Table~\ref{table1}.

For the A220 dataset, the best-fit value of $a$ is equal to $0.54 \pm 0.02$ with an intrinsic scatter of 28\%. The best-fit value of $a$ is consistent with that obtained in G22 to within $1\sigma$.
This scatter is smaller than the scatter of around  45\%  obtained  in G22 (for $z<0.9$) and  ~\cite{Khadka21}, or the value of 50\% obtained in ~\cite{Liu22}. 

For the D17 dataset, we find $a= 0.4 \pm 0.03$ with an intrinsic scatter of 35\%. The best-fit value of $a$ is again consistent with that obtained in G22 to within $1\sigma$. 
The scatter obtained in this work, however,  is substantially  larger  than the value of 15\% obtained in G22 (for $z<0.9$) or the value of 26\% obtained using quasars in ~\cite{Dai21} (for $z\leq 1$). 


Therefore, we find that although the Amati  slope is consistent with previous results, which used disparate external observables as distance anchors,
the intrinsic scatter using the HII galaxies as distance anchors is still very large. Therefore, the Amati relation obtained using external datasets as anchors, especially at high redshifts to break the circularity, may not always be robust  for precision Cosmology. We should also point out that another caveat in using HII galaxy dataset as distance anchors is that there could be an evolution in the $L-\sigma$  between the low- and high-redshift samples~\cite{CaoRatra23}. This could also explain the large scatter, which we have observed in the Amati relation for the two datasets. 

Recent analyses of the Amati relation have found much larger scatter compared to the two datasets we have analyzed. A scatter of 39\% was obtained from a joint fitting of the  cosmology and Amati relation~\cite{CaoRatra24}. A subsequent analysis by the same group found a scatter of $55\%-65\%$ using an updated GRB dataset~\cite{CaoRatra25}.
Using cosmic chronometers as distance anchors for a sample of long GRBs from Fermi-GRB, an intrinsic  scatter of 50-65\% was obtained for $z \leq 1.4$~\cite{Wangchronometer} and 41-59\% for $z<1.4$ or $z<1.965$~\cite{Huangchrono}. Furthermore, one should also select a GRB dataset, which is standardizable~\cite{Khadka20,Khadka21,CaoRatra24,CaoRatra25,Wangchronometer,Huangchrono} and independent of the underlying cosmological model, in order to make it suitable for cosmological analysis.

\begin{table}[h]
\caption{\label{table1}Summary of our results for the Amati relation for A220 and D17  GRB datasets for $z<2.6$. Both datasets show a high intrinsic scatter in the Amati relation.}
\begin{tabular}{ |c| c|c| c| }
\hline
\textbf{Dataset} & \textbf{a} &\textbf{b} & \textbf{$\sigma_s$} \\
\hline
A220 &$0.54 \pm 0.02$ &$-25.42 \pm 1.2$ & $0.28 \pm 0.013$ \\
D17 &  $0.40 \pm 0.03$ &$-18.26 \pm 1.8 $ &$0.35 \pm 0.03$\\
\hline
\end{tabular}
\end{table}

\section{Conclusions}
\label{sec:conclusions}
In this work, we have used  186 HII galaxy distances as anchors to calibrate the efficacy of the Amati relation between $E_p$ and $E_{iso}$. We first  used ANN based interpolation to reconstruct $D_L$ for any GRB redshift.
These interpolated values of $D_L$ were used to  obtain $E_{iso}$ for the redshifts corresponding to  two  different GRB datasets (A220 and D17), which have previously been used for analysis of the Amati relation in a number of works, including G22.

 We find that the best-fit parameters for both the datasets   are consistent with those obtained G22 to within $1\sigma$.   \rthis{However, there is about a 3$\sigma$ discrepancy in the best-fit values for the slope and intercept between the two datasets. One possible reason for the same could be that the A220 GRB dataset has been standardized, whereas the D17 dataset has analyzed using Type Ia Supernovae as external anchors.}
 Furthermore, the intrinsic scatter obtained for A220 and D17  is quite large, viz. 28\% and 35\%,  respectively.
 Although this scatter is roughly  comparable to the values  obtained using other observables as distance anchors, it implies  that the Amati relation using HII galaxies as distance anchors  cannot be used for precision Cosmology. 

 Looking ahead, GRB observations from  SVOM (launched in June 2024)~\cite{SVOM} and upcoming satellites such as THESEUS~\cite{Theseus} are expected to provide further constraints on the Amati relation.
 
\vspace{6pt}

\dataavailability{ No new data were generated for this article. The GRB dataset used for this analysis can be found in ~\cite{Khadka21} and ~\cite{Demianski17}. The HII galaxy dataset can be found in the references in ~\cite{Melia25}. The datasets have also been uploaded to \url{https://github.com/quantumnaan/amati_test}. }

\acknowledgments{RO would like to thank the JST Sakura Science Program (Grant No. M2025L0418003) and the JASSO Scholarship for supporting his travel to and stay at IIT Hyderabad. We also thank Yuva Himanshu Pallam for helping us to collate the HII galaxy dataset. We are grateful to the anonymous referees for several useful and constructive comments on our manuscript. }

\conflictsofinterest{The authors declare no conflicts of interest.} 

\begin{adjustwidth}{-\extralength}{0cm}

\reftitle{References}



\bibliography{main}

\begin{thebibliography}{999}

\bibitem[{Kumar} and {Zhang}(2015)]{Kumar}
{Kumar}, P.; {Zhang}, B.
\newblock {The physics of gamma-ray bursts \& relativistic jets}.
\newblock {\em \physrep} {\bf 2015}, {\em 561},~1--109,  \href{http://arxiv.org/abs/1410.0679}{{\normalfont [arXiv:astro-ph.HE/1410.0679]}}.
\newblock {\url{https://doi.org/10.1016/j.physrep.2014.09.008}}.

\bibitem[{Kouveliotou} et~al.(1993){Kouveliotou}, {Meegan}, {Fishman}, {Bhat}, {Briggs}, {Koshut}, {Paciesas}, and {Pendleton}]{Kouv93}
{Kouveliotou}, C.; {Meegan}, C.A.; {Fishman}, G.J.; {Bhat}, N.P.; {Briggs}, M.S.; {Koshut}, T.M.; {Paciesas}, W.S.; {Pendleton}, G.N.
\newblock {Identification of Two Classes of Gamma-Ray Bursts}.
\newblock {\em \apjl} {\bf 1993}, {\em 413},~L101.
\newblock {\url{https://doi.org/10.1086/186969}}.

\bibitem[{Woosley} and {Bloom}(2006)]{Bloom}
{Woosley}, S.E.; {Bloom}, J.S.
\newblock {The Supernova Gamma-Ray Burst Connection}.
\newblock {\em \araa} {\bf 2006}, {\em 44},~507--556,  \href{http://arxiv.org/abs/astro-ph/0609142}{{\normalfont [arXiv:astro-ph/astro-ph/0609142]}}.
\newblock {\url{https://doi.org/10.1146/annurev.astro.43.072103.150558}}.

\bibitem[{Nakar}(2007)]{Nakar}
{Nakar}, E.
\newblock {Short-hard gamma-ray bursts}.
\newblock {\em \physrep} {\bf 2007}, {\em 442},~166--236,  \href{http://arxiv.org/abs/astro-ph/0701748}{{\normalfont [arXiv:astro-ph/astro-ph/0701748]}}.
\newblock {\url{https://doi.org/10.1016/j.physrep.2007.02.005}}.

\bibitem[{Tsvetkova} et~al.(2025){Tsvetkova}, {Amati}, {Bulla}, {Burderi}, {Frederiks}, {Frontera}, {Guidorzi}, {Riggio}, {di Salvo}, {Sanna}, and {Sviridov}]{Anastasia}
{Tsvetkova}, A.; {Amati}, L.; {Bulla}, M.; {Burderi}, L.; {Frederiks}, D.; {Frontera}, F.; {Guidorzi}, C.; {Riggio}, A.; {di Salvo}, T.; {Sanna}, A.;  et~al.
\newblock {Gamma-ray burst taxonomy: Looking for the third class on the spectral peak energy-duration plane in the rest frame}.
\newblock {\em \aap} {\bf 2025}, {\em 698},~A169,  \href{http://arxiv.org/abs/2504.20263}{{\normalfont [arXiv:astro-ph.HE/2504.20263]}}.
\newblock {\url{https://doi.org/10.1051/0004-6361/202452673}}.

\bibitem[{Parsotan} and {Ito}(2022)]{Ito}
{Parsotan}, T.; {Ito}, H.
\newblock {GRB Prompt Emission: Observed Correlations and Their Interpretations}.
\newblock {\em Universe} {\bf 2022}, {\em 8},~310,  \href{http://arxiv.org/abs/2204.09729}{{\normalfont [arXiv:astro-ph.HE/2204.09729]}}.
\newblock {\url{https://doi.org/10.3390/universe8060310}}.

\bibitem[{Dainotti} and {Del Vecchio}(2017)]{Delvecchio}
{Dainotti}, M.G.; {Del Vecchio}, R.
\newblock {Gamma Ray Burst afterglow and prompt-afterglow relations: An overview}.
\newblock {\em \nar} {\bf 2017}, {\em 77},~23--61,  \href{http://arxiv.org/abs/1703.06876}{{\normalfont [arXiv:astro-ph.HE/1703.06876]}}.
\newblock {\url{https://doi.org/10.1016/j.newar.2017.04.001}}.

\bibitem[{Dainotti} and {Amati}(2018)]{DainottiAmati}
{Dainotti}, M.G.; {Amati}, L.
\newblock {Gamma-ray Burst Prompt Correlations: Selection and Instrumental Effects}.
\newblock {\em \pasp} {\bf 2018}, {\em 130},~051001,  \href{http://arxiv.org/abs/1704.00844}{{\normalfont [arXiv:astro-ph.HE/1704.00844]}}.
\newblock {\url{https://doi.org/10.1088/1538-3873/aaa8d7}}.

\bibitem[{Moresco} et~al.(2022){Moresco}, {Amati}, {Amendola}, {Birrer}, {Blakeslee}, {Cantiello}, {Cimatti}, {Darling}, {Della Valle}, {Fishbach}, {Grillo}, {Hamaus}, {Holz}, {Izzo}, {Jimenez}, {Lusso}, {Meneghetti}, {Piedipalumbo}, {Pisani}, {Pourtsidou}, {Pozzetti}, {Quartin}, {Risaliti}, {Rosati}, and {Verde}]{Moresco}
{Moresco}, M.; {Amati}, L.; {Amendola}, L.; {Birrer}, S.; {Blakeslee}, J.P.; {Cantiello}, M.; {Cimatti}, A.; {Darling}, J.; {Della Valle}, M.; {Fishbach}, M.;  et~al.
\newblock {Unveiling the Universe with emerging cosmological probes}.
\newblock {\em Living Reviews in Relativity} {\bf 2022}, {\em 25},~6,  \href{http://arxiv.org/abs/2201.07241}{{\normalfont [arXiv:astro-ph.CO/2201.07241]}}.
\newblock {\url{https://doi.org/10.1007/s41114-022-00040-z}}.

\bibitem[{Luongo} and {Muccino}(2021)]{Luongo21}
{Luongo}, O.; {Muccino}, M.
\newblock {A Roadmap to Gamma-Ray Bursts: New Developments and Applications to Cosmology}.
\newblock {\em Galaxies} {\bf 2021}, {\em 9},~77,  \href{http://arxiv.org/abs/2110.14408}{{\normalfont [arXiv:astro-ph.HE/2110.14408]}}.
\newblock {\url{https://doi.org/10.3390/galaxies9040077}}.

\bibitem[{Dainotti} et~al.(2022){Dainotti}, {Nielson}, {Sarracino}, {Rinaldi}, {Nagataki}, {Capozziello}, {Gnedin}, and {Bargiacchi}]{Dainotti22}
{Dainotti}, M.G.; {Nielson}, V.; {Sarracino}, G.; {Rinaldi}, E.; {Nagataki}, S.; {Capozziello}, S.; {Gnedin}, O.Y.; {Bargiacchi}, G.
\newblock {Optical and X-ray GRB Fundamental Planes as cosmological distance indicators}.
\newblock {\em \mnras} {\bf 2022}, {\em 514},~1828--1856,  \href{http://arxiv.org/abs/2203.15538}{{\normalfont [arXiv:astro-ph.CO/2203.15538]}}.
\newblock {\url{https://doi.org/10.1093/mnras/stac1141}}.

\bibitem[{Bargiacchi} et~al.(2025){Bargiacchi}, {Dainotti}, and {Capozziello}]{Bargiacchi}
{Bargiacchi}, G.; {Dainotti}, M.G.; {Capozziello}, S.
\newblock {High-redshift cosmology by Gamma-Ray Bursts: An overview}.
\newblock {\em \nar} {\bf 2025}, {\em 100},~101712,  \href{http://arxiv.org/abs/2408.10707}{{\normalfont [arXiv:astro-ph.CO/2408.10707]}}.
\newblock {\url{https://doi.org/10.1016/j.newar.2024.101712}}.

\bibitem[{Amati} et~al.(2002){Amati}, {Frontera}, {Tavani}, {in't Zand}, {Antonelli}, {Costa}, {Feroci}, {Guidorzi}, {Heise}, {Masetti}, {Montanari}, {Nicastro}, {Palazzi}, {Pian}, {Piro}, and {Soffitta}]{Amati02}
{Amati}, L.; {Frontera}, F.; {Tavani}, M.; {in't Zand}, J.J.M.; {Antonelli}, A.; {Costa}, E.; {Feroci}, M.; {Guidorzi}, C.; {Heise}, J.; {Masetti}, N.;  et~al.
\newblock {Intrinsic spectra and energetics of BeppoSAX Gamma-Ray Bursts with known redshifts}.
\newblock {\em \aap} {\bf 2002}, {\em 390},~81--89,  \href{http://arxiv.org/abs/astro-ph/0205230}{{\normalfont [arXiv:astro-ph/astro-ph/0205230]}}.
\newblock {\url{https://doi.org/10.1051/0004-6361:20020722}}.

\bibitem[{Amati}(2006)]{Amati06}
{Amati}, L.
\newblock {The E$_{p,i}$-E$_{iso}$ correlation in gamma-ray bursts: updated observational status, re-analysis and main implications}.
\newblock {\em \mnras} {\bf 2006}, {\em 372},~233--245,  \href{http://arxiv.org/abs/astro-ph/0601553}{{\normalfont [arXiv:astro-ph/astro-ph/0601553]}}.
\newblock {\url{https://doi.org/10.1111/j.1365-2966.2006.10840.x}}.

\bibitem[{Collazzi} et~al.(2012){Collazzi}, {Schaefer}, {Goldstein}, and {Preece}]{Collazzi}
{Collazzi}, A.C.; {Schaefer}, B.E.; {Goldstein}, A.; {Preece}, R.D.
\newblock {A Significant Problem with Using the Amati Relation for Cosmological Purposes}.
\newblock {\em \apj} {\bf 2012}, {\em 747},~39,  \href{http://arxiv.org/abs/1112.4347}{{\normalfont [arXiv:astro-ph.HE/1112.4347]}}.
\newblock {\url{https://doi.org/10.1088/0004-637X/747/1/39}}.

\bibitem[{Amati} et~al.(2008){Amati}, {Guidorzi}, {Frontera}, {Della Valle}, {Finelli}, {Landi}, and {Montanari}]{Amati08}
{Amati}, L.; {Guidorzi}, C.; {Frontera}, F.; {Della Valle}, M.; {Finelli}, F.; {Landi}, R.; {Montanari}, E.
\newblock {Measuring the cosmological parameters with the E$_{p,i}$-E$_{iso}$ correlation of gamma-ray bursts}.
\newblock {\em \mnras} {\bf 2008}, {\em 391},~577--584,  \href{http://arxiv.org/abs/0805.0377}{{\normalfont [arXiv:astro-ph/0805.0377]}}.
\newblock {\url{https://doi.org/10.1111/j.1365-2966.2008.13943.x}}.

\bibitem[{Khadka} and {Ratra}(2020)]{Khadka20}
{Khadka}, N.; {Ratra}, B.
\newblock {Constraints on cosmological parameters from gamma-ray burst peak photon energy and bolometric fluence measurements and other data}.
\newblock {\em \mnras} {\bf 2020}, {\em 499},~391--403,  \href{http://arxiv.org/abs/2007.13907}{{\normalfont [arXiv:astro-ph.CO/2007.13907]}}.
\newblock {\url{https://doi.org/10.1093/mnras/staa2779}}.

\bibitem[{Khadka} et~al.(2021){Khadka}, {Luongo}, {Muccino}, and {Ratra}]{Khadka21}
{Khadka}, N.; {Luongo}, O.; {Muccino}, M.; {Ratra}, B.
\newblock {Do gamma-ray burst measurements provide a useful test of cosmological models?}
\newblock {\em \jcap} {\bf 2021}, {\em 2021},~042,  \href{http://arxiv.org/abs/2105.12692}{{\normalfont [arXiv:astro-ph.CO/2105.12692]}}.
\newblock {\url{https://doi.org/10.1088/1475-7516/2021/09/042}}.

\bibitem[{Cao} and {Ratra}(2024)]{CaoRatra24}
{Cao}, S.; {Ratra}, B.
\newblock {Testing the standardizability of, and deriving cosmological constraints from, a new Amati-correlated gamma-ray burst data compilation}.
\newblock {\em \jcap} {\bf 2024}, {\em 2024},~093,  \href{http://arxiv.org/abs/2404.08697}{{\normalfont [arXiv:astro-ph.CO/2404.08697]}}.
\newblock {\url{https://doi.org/10.1088/1475-7516/2024/10/093}}.

\bibitem[{Cao} and {Ratra}(2025)]{CaoRatra25}
{Cao}, S.; {Ratra}, B.
\newblock {Testing the consistency of new Amati-correlated gamma-ray burst dataset cosmological constraints with those from better-established cosmological data}.
\newblock {\em \jcap} {\bf 2025}, {\em 2025},~081,  \href{http://arxiv.org/abs/2502.08429}{{\normalfont [arXiv:astro-ph.CO/2502.08429]}}.
\newblock {\url{https://doi.org/10.1088/1475-7516/2025/09/081}}.

\bibitem[{Liang} et~al.(2008){Liang}, {Xiao}, {Liu}, and {Zhang}]{LiangAmati}
{Liang}, N.; {Xiao}, W.K.; {Liu}, Y.; {Zhang}, S.N.
\newblock {A Cosmology-Independent Calibration of Gamma-Ray Burst Luminosity Relations and the Hubble Diagram}.
\newblock {\em \apj} {\bf 2008}, {\em 685},~354--360,  \href{http://arxiv.org/abs/0802.4262}{{\normalfont [arXiv:astro-ph/0802.4262]}}.
\newblock {\url{https://doi.org/10.1086/590903}}.

\bibitem[{Liang} et~al.(2022){Liang}, {Li}, {Xie}, and {Wu}]{Liang22}
{Liang}, N.; {Li}, Z.; {Xie}, X.; {Wu}, P.
\newblock {Calibrating Gamma-Ray Bursts by Using a Gaussian Process with Type Ia Supernovae}.
\newblock {\em \apj} {\bf 2022}, {\em 941},~84,  \href{http://arxiv.org/abs/2211.02473}{{\normalfont [arXiv:astro-ph.CO/2211.02473]}}.
\newblock {\url{https://doi.org/10.3847/1538-4357/aca08a}}.

\bibitem[{Kodama} et~al.(2008){Kodama}, {Yonetoku}, {Murakami}, {Tanabe}, {Tsutsui}, and {Nakamura}]{Kodama08}
{Kodama}, Y.; {Yonetoku}, D.; {Murakami}, T.; {Tanabe}, S.; {Tsutsui}, R.; {Nakamura}, T.
\newblock {Gamma-ray bursts in 1.8 < z < 5.6 suggest that the time variation of the dark energy is small}.
\newblock {\em \mnras} {\bf 2008}, {\em 391},~L1--L4,  \href{http://arxiv.org/abs/0802.3428}{{\normalfont [arXiv:astro-ph/0802.3428]}}.
\newblock {\url{https://doi.org/10.1111/j.1745-3933.2008.00508.x}}.

\bibitem[{Demianski} et~al.(2017){Demianski}, {Piedipalumbo}, {Sawant}, and {Amati}]{Demianski17}
{Demianski}, M.; {Piedipalumbo}, E.; {Sawant}, D.; {Amati}, L.
\newblock {Cosmology with gamma-ray bursts. I. The Hubble diagram through the calibrated E$_{p,I}$-E$_{iso}$ correlation}.
\newblock {\em \aap} {\bf 2017}, {\em 598},~A112,  \href{http://arxiv.org/abs/1610.00854}{{\normalfont [arXiv:astro-ph.CO/1610.00854]}}.
\newblock {\url{https://doi.org/10.1051/0004-6361/201628909}}.

\bibitem[{Liu} et~al.(2022){Liu}, {Liang}, {Xie}, {Yuan}, {Yu}, and {Wu}]{Liu22copula}
{Liu}, Y.; {Liang}, N.; {Xie}, X.; {Yuan}, Z.; {Yu}, H.; {Wu}, P.
\newblock {Gamma-Ray Burst Constraints on Cosmological Models from the Improved Amati Correlation}.
\newblock {\em \apj} {\bf 2022}, {\em 935},~7,  \href{http://arxiv.org/abs/2207.00455}{{\normalfont [arXiv:astro-ph.CO/2207.00455]}}.
\newblock {\url{https://doi.org/10.3847/1538-4357/ac7de5}}.

\bibitem[{Huang} et~al.(2025){Huang}, {Luo}, {Zhang}, {Feng}, {Wu}, {Liu}, and {Liang}]{Huang25}
{Huang}, Z.; {Luo}, X.; {Zhang}, B.; {Feng}, J.; {Wu}, P.; {Liu}, Y.; {Liang}, N.
\newblock {Gamma-Ray Bursts Calibrated by Using Artificial Neural Networks from the Pantheon+ Sample}.
\newblock {\em Universe} {\bf 2025}, {\em 11},~241,  \href{http://arxiv.org/abs/2506.08929}{{\normalfont [arXiv:astro-ph.CO/2506.08929]}}.
\newblock {\url{https://doi.org/10.3390/universe11080241}}.

\bibitem[{Montiel} et~al.(2021){Montiel}, {Cabrera}, and {Hidalgo}]{Montiel}
{Montiel}, A.; {Cabrera}, J.I.; {Hidalgo}, J.C.
\newblock {Improving sampling and calibration of gamma-ray bursts as distance indicators}.
\newblock {\em \mnras} {\bf 2021}, {\em 501},~3515--3526,  \href{http://arxiv.org/abs/2003.03387}{{\normalfont [arXiv:astro-ph.HE/2003.03387]}}.
\newblock {\url{https://doi.org/10.1093/mnras/staa3926}}.

\bibitem[{Amati} et~al.(2019){Amati}, {D'Agostino}, {Luongo}, {Muccino}, and {Tantalo}]{Amati19}
{Amati}, L.; {D'Agostino}, R.; {Luongo}, O.; {Muccino}, M.; {Tantalo}, M.
\newblock {Addressing the circularity problem in the E$_{p}$-E$_{iso}$ correlation of gamma-ray bursts}.
\newblock {\em \mnras} {\bf 2019}, {\em 486},~L46--L51,  \href{http://arxiv.org/abs/1811.08934}{{\normalfont [arXiv:astro-ph.HE/1811.08934]}}.
\newblock {\url{https://doi.org/10.1093/mnrasl/slz056}}.

\bibitem[{Luongo} and {Muccino}(2021)]{LuongoOHD}
{Luongo}, O.; {Muccino}, M.
\newblock {Model-independent calibrations of gamma-ray bursts using machine learning}.
\newblock {\em \mnras} {\bf 2021}, {\em 503},~4581--4600,  \href{http://arxiv.org/abs/2011.13590}{{\normalfont [arXiv:astro-ph.CO/2011.13590]}}.
\newblock {\url{https://doi.org/10.1093/mnras/stab795}}.

\bibitem[{Luongo} and {Muccino}(2022)]{Luongo}
{Luongo}, O.; {Muccino}, M.
\newblock {Intermediate redshift calibration of Gamma-ray Bursts and cosmic constraints in non-flat cosmology}.
\newblock {\em arXiv e-prints} {\bf 2022}, p. arXiv:2207.00440,  \href{http://arxiv.org/abs/2207.00440}{{\normalfont [arXiv:astro-ph.CO/2207.00440]}}.

\bibitem[{Kumar} et~al.(2023){Kumar}, {Rani}, {Jain}, {Mahajan}, and {Mukherjee}]{Jain23}
{Kumar}, D.; {Rani}, N.; {Jain}, D.; {Mahajan}, S.; {Mukherjee}, A.
\newblock {Gamma rays bursts: a viable cosmological probe?}
\newblock {\em \jcap} {\bf 2023}, {\em 2023},~021,  \href{http://arxiv.org/abs/2212.05731}{{\normalfont [arXiv:astro-ph.CO/2212.05731]}}.
\newblock {\url{https://doi.org/10.1088/1475-7516/2023/07/021}}.

\bibitem[{Wang} and {Liang}(2024)]{Wangchronometer}
{Wang}, H.; {Liang}, N.
\newblock {Constraints from Fermi observations of long gamma-ray bursts on cosmological parameters}.
\newblock {\em \mnras} {\bf 2024}, {\em 533},~743--755,  \href{http://arxiv.org/abs/2405.14357}{{\normalfont [arXiv:astro-ph.CO/2405.14357]}}.
\newblock {\url{https://doi.org/10.1093/mnras/stae1825}}.

\bibitem[{Huang} et~al.(2025){Huang}, {Xiong}, {Luo}, {Wang}, {Liu}, and {Liang}]{Huangchrono}
{Huang}, Z.; {Xiong}, Z.; {Luo}, X.; {Wang}, G.; {Liu}, Y.; {Liang}, N.
\newblock {Gamma-ray bursts calibrated from the observational H(z) data in artificial neural network framework}.
\newblock {\em Journal of High Energy Astrophysics} {\bf 2025}, {\em 47},~100377,  \href{http://arxiv.org/abs/2502.10037}{{\normalfont [arXiv:astro-ph.CO/2502.10037]}}.
\newblock {\url{https://doi.org/10.1016/j.jheap.2025.100377}}.

\bibitem[{Govindaraj} and {Desai}(2022)]{Gowri}
{Govindaraj}, G.; {Desai}, S.
\newblock {Low redshift calibration of the Amati relation using galaxy clusters}.
\newblock {\em \jcap} {\bf 2022}, {\em 2022},~069,  \href{http://arxiv.org/abs/2208.00895}{{\normalfont [arXiv:astro-ph.HE/2208.00895]}}.
\newblock {\url{https://doi.org/10.1088/1475-7516/2022/10/069}}.

\bibitem[{Dai} et~al.(2021){Dai}, {Zheng}, {Li}, {Gao}, and {Zhu}]{Dai21}
{Dai}, Y.; {Zheng}, X.G.; {Li}, Z.X.; {Gao}, H.; {Zhu}, Z.H.
\newblock {Redshift evolution of the Amati relation: Calibrated results from the Hubble diagram of quasars at high redshifts}.
\newblock {\em \aap} {\bf 2021}, {\em 651},~L8,  \href{http://arxiv.org/abs/2111.05544}{{\normalfont [arXiv:astro-ph.HE/2111.05544]}}.
\newblock {\url{https://doi.org/10.1051/0004-6361/202140895}}.

\bibitem[{Terlevich} and {Melnick}(1981)]{Terlevich81}
{Terlevich}, R.; {Melnick}, J.
\newblock {The dynamics and chemical composition of giant extragalactic H II regions}.
\newblock {\em \mnras} {\bf 1981}, {\em 195},~839--851.
\newblock {\url{https://doi.org/10.1093/mnras/195.4.839}}.

\bibitem[{Siegel} et~al.(2005){Siegel}, {Guzm{\'a}n}, {Gallego}, {Ordu{\~n}a L{\'o}pez}, and {Rodr{\'\i}guez Hidalgo}]{Siegel05}
{Siegel}, E.R.; {Guzm{\'a}n}, R.; {Gallego}, J.P.; {Ordu{\~n}a L{\'o}pez}, M.; {Rodr{\'\i}guez Hidalgo}, P.
\newblock {Towards a precision cosmology from starburst galaxies at z > 2}.
\newblock {\em \mnras} {\bf 2005}, {\em 356},~1117--1122,  \href{http://arxiv.org/abs/astro-ph/0410612}{{\normalfont [arXiv:astro-ph/astro-ph/0410612]}}.
\newblock {\url{https://doi.org/10.1111/j.1365-2966.2004.08539.x}}.

\bibitem[{Ch{\'a}vez} et~al.(2012){Ch{\'a}vez}, {Terlevich}, {Terlevich}, {Plionis}, {Bresolin}, {Basilakos}, and {Melnick}]{Chavez12}
{Ch{\'a}vez}, R.; {Terlevich}, E.; {Terlevich}, R.; {Plionis}, M.; {Bresolin}, F.; {Basilakos}, S.; {Melnick}, J.
\newblock {Determining the Hubble constant using giant extragalactic H II regions and H II galaxies}.
\newblock {\em \mnras} {\bf 2012}, {\em 425},~L56--L60,  \href{http://arxiv.org/abs/1203.6222}{{\normalfont [arXiv:astro-ph.CO/1203.6222]}}.
\newblock {\url{https://doi.org/10.1111/j.1745-3933.2012.01299.x}}.

\bibitem[{Mania} and {Ratra}(2012)]{Mania12}
{Mania}, D.; {Ratra}, B.
\newblock {Constraints on dark energy from H II starburst galaxy apparent magnitude versus redshift data}.
\newblock {\em Physics Letters B} {\bf 2012}, {\em 715},~9--14,  \href{http://arxiv.org/abs/1110.5626}{{\normalfont [arXiv:astro-ph.CO/1110.5626]}}.
\newblock {\url{https://doi.org/10.1016/j.physletb.2012.07.011}}.

\bibitem[{Wei} and {Melia}(2025)]{Melia25}
{Wei}, J.J.; {Melia}, F.
\newblock {Model selection using the HII galaxy Hubble diagram}.
\newblock {\em \mnras} {\bf 2025}, {\em 542},~L19--L23,  \href{http://arxiv.org/abs/2506.04819}{{\normalfont [arXiv:astro-ph.CO/2506.04819]}}.
\newblock {\url{https://doi.org/10.1093/mnrasl/slaf060}}.

\bibitem[{Zheng} et~al.(2025){Zheng}, {Qiang}, {You}, and {Kumar}]{Zheng25}
{Zheng}, J.; {Qiang}, D.C.; {You}, Z.Q.; {Kumar}, D.
\newblock {Quantifying the Impact of 2D and 3D BAO Measurements on the Cosmic Distance Duality Relation with HII Galaxy observation}.
\newblock {\em arXiv e-prints} {\bf 2025}, p. arXiv:2507.17113,  \href{http://arxiv.org/abs/2507.17113}{{\normalfont [arXiv:astro-ph.CO/2507.17113]}}.
\newblock {\url{https://doi.org/10.48550/arXiv.2507.17113}}.

\bibitem[{Demianski} et~al.(2017){Demianski}, {Piedipalumbo}, {Sawant}, and {Amati}]{Demianski2}
{Demianski}, M.; {Piedipalumbo}, E.; {Sawant}, D.; {Amati}, L.
\newblock {Cosmology with gamma-ray bursts. II. Cosmography challenges and cosmological scenarios for the accelerated Universe}.
\newblock {\em \aap} {\bf 2017}, {\em 598},~A113,  \href{http://arxiv.org/abs/1609.09631}{{\normalfont [arXiv:astro-ph.CO/1609.09631]}}.
\newblock {\url{https://doi.org/10.1051/0004-6361/201628911}}.

\bibitem[{Gonz{\'a}lez-Mor{\'a}n} et~al.(2021){Gonz{\'a}lez-Mor{\'a}n}, {Ch{\'a}vez}, {Terlevich}, {Terlevich}, {Fern{\'a}ndez-Arenas}, {Bresolin}, {Plionis}, {Melnick}, {Basilakos}, and {Telles}]{gonzalez}
{Gonz{\'a}lez-Mor{\'a}n}, A.L.; {Ch{\'a}vez}, R.; {Terlevich}, E.; {Terlevich}, R.; {Fern{\'a}ndez-Arenas}, D.; {Bresolin}, F.; {Plionis}, M.; {Melnick}, J.; {Basilakos}, S.; {Telles}, E.
\newblock {Independent cosmological constraints from high-z H II galaxies: new results from VLT-KMOS data}.
\newblock {\em \mnras} {\bf 2021}, {\em 505},~1441--1457,  \href{http://arxiv.org/abs/2105.04025}{{\normalfont [arXiv:astro-ph.CO/2105.04025]}}.
\newblock {\url{https://doi.org/10.1093/mnras/stab1385}}.

\bibitem[{Llerena} et~al.(2023){Llerena}, {Amor{\'\i}n}, {Pentericci}, {Calabr{\`o}}, {Shapley}, {Boutsia}, {P{\'e}rez-Montero}, {V{\'\i}lchez}, and {Nakajima}]{Llerena23}
{Llerena}, M.; {Amor{\'\i}n}, R.; {Pentericci}, L.; {Calabr{\`o}}, A.; {Shapley}, A.E.; {Boutsia}, K.; {P{\'e}rez-Montero}, E.; {V{\'\i}lchez}, J.M.; {Nakajima}, K.
\newblock {Ionized gas kinematics and chemical abundances of low-mass star-forming galaxies at z {\ensuremath{\sim}} 3}.
\newblock {\em \aap} {\bf 2023}, {\em 676},~A53,  \href{http://arxiv.org/abs/2303.01536}{{\normalfont [arXiv:astro-ph.GA/2303.01536]}}.
\newblock {\url{https://doi.org/10.1051/0004-6361/202346232}}.

\bibitem[{de Graaff} et~al.(2024){de Graaff}, {Rix}, {Carniani}, {Suess}, {Charlot}, {Curtis-Lake}, {Arribas}, {Baker}, {Boyett}, {Bunker}, {Cameron}, {Chevallard}, {Curti}, {Eisenstein}, {Franx}, {Hainline}, {Hausen}, {Ji}, {Johnson}, {Jones}, {Maiolino}, {Maseda}, {Nelson}, {Parlanti}, {Rawle}, {Robertson}, {Tacchella}, {{\"U}bler}, {Williams}, {Willmer}, and {Willott}]{DeGraaf}
{de Graaff}, A.; {Rix}, H.W.; {Carniani}, S.; {Suess}, K.A.; {Charlot}, S.; {Curtis-Lake}, E.; {Arribas}, S.; {Baker}, W.M.; {Boyett}, K.; {Bunker}, A.J.;  et~al.
\newblock {Ionised gas kinematics and dynamical masses of z {\ensuremath{\gtrsim}} 6 galaxies from JADES/NIRSpec high-resolution spectroscopy}.
\newblock {\em \aap} {\bf 2024}, {\em 684},~A87,  \href{http://arxiv.org/abs/2308.09742}{{\normalfont [arXiv:astro-ph.GA/2308.09742]}}.
\newblock {\url{https://doi.org/10.1051/0004-6361/202347755}}.

\bibitem[{Ball} and {Brunner}(2010)]{Ball10}
{Ball}, N.M.; {Brunner}, R.J.
\newblock {Data Mining and Machine Learning in Astronomy}.
\newblock {\em International Journal of Modern Physics D} {\bf 2010}, {\em 19},~1049--1106,  \href{http://arxiv.org/abs/0906.2173}{{\normalfont [arXiv:astro-ph.IM/0906.2173]}}.
\newblock {\url{https://doi.org/10.1142/S0218271810017160}}.

\bibitem[Efron and Tibshirani(1993)]{Efron1993bootstrap}
Efron, B.; Tibshirani, R.J.
\newblock {\em An introduction to the bootstrap}; Chapman \& Hall/CRC,  1993.

\bibitem[{Cao} and {Ratra}(2024)]{CaoRatra23}
{Cao}, S.; {Ratra}, B.
\newblock {Low- and high-redshift H II starburst galaxies obey different luminosity-velocity dispersion relations}.
\newblock {\em \prd} {\bf 2024}, {\em 109},~123527,  \href{http://arxiv.org/abs/2310.15812}{{\normalfont [arXiv:astro-ph.CO/2310.15812]}}.
\newblock {\url{https://doi.org/10.1103/PhysRevD.109.123527}}.

\bibitem[Foreman-Mackey et~al.(2013)Foreman-Mackey, Hogg, Lang, and Goodman]{emcee}
Foreman-Mackey, D.; Hogg, D.W.; Lang, D.; Goodman, J.
\newblock {emcee: The MCMC Hammer}.
\newblock {\em Publ. Astron. Soc. Pac.} {\bf 2013}, {\em 125},~306--312,  \href{http://arxiv.org/abs/1202.3665}{{\normalfont [arXiv:astro-ph.IM/1202.3665]}}.
\newblock {\url{https://doi.org/10.1086/670067}}.

\bibitem[{Liu} et~al.(2022){Liu}, {Chen}, {Liang}, {Yuan}, {Yu}, and {Wu}]{Liu22}
{Liu}, Y.; {Chen}, F.; {Liang}, N.; {Yuan}, Z.; {Yu}, H.; {Wu}, P.
\newblock {The Improved Amati Correlations from Gaussian Copula}.
\newblock {\em \apj} {\bf 2022}, {\em 931},~50,  \href{http://arxiv.org/abs/2203.03178}{{\normalfont [arXiv:astro-ph.CO/2203.03178]}}.
\newblock {\url{https://doi.org/10.3847/1538-4357/ac66d3}}.

\bibitem[{Bernardini} et~al.(2021){Bernardini}, {Cordier}, and {Wei}]{SVOM}
{Bernardini}, M.G.; {Cordier}, B.; {Wei}, J.
\newblock {The SVOM Mission}.
\newblock {\em Galaxies} {\bf 2021}, {\em 9},~113.
\newblock {\url{https://doi.org/10.3390/galaxies9040113}}.

\bibitem[{Amati} et~al.(2018){Amati}, {O'Brien}, {G{\"o}tz}, {Bozzo}, {Tenzer}, {Frontera}, {Ghirlanda}, {Labanti}, {Osborne}, {Stratta}, {Tanvir}, {Willingale}, {Attina}, {Campana}, {Castro-Tirado}, {Contini}, {Fuschino}, {Gomboc}, {Hudec}, {Orleanski}, {Renotte}, {Rodic}, {Bagoly}, {Blain}, {Callanan}, {Covino}, {Ferrara}, {Le Floch}, {Marisaldi}, {Mereghetti}, {Rosati}, {Vacchi}, {D'Avanzo}, {Giommi}, {Piranomonte}, {Piro}, {Reglero}, {Rossi}, {Santangelo}, {Salvaterra}, {Tagliaferri}, {Vergani}, {Vinciguerra}, {Briggs}, {Campolongo}, {Ciolfi}, {Connaughton}, {Cordier}, {Morelli}, {Orlandini}, {Adami}, {Argan}, {Atteia}, {Auricchio}, {Balazs}, {Baldazzi}, {Basa}, {Basak}, {Bellutti}, {Bernardini}, {Bertuccio}, {Braga}, {Branchesi}, {Brandt}, {Brocato}, {Budtz-Jorgensen}, {Bulgarelli}, {Burderi}, {Camp}, {Capozziello}, {Caruana}, {Casella}, {Cenko}, {Chardonnet}, {Ciardi}, {Colafrancesco}, {Dainotti}, {D'Elia}, {De Martino}, {De Pasquale}, {Del Monte}, {Della Valle}, {Drago}, {Evangelista}, {Feroci},
  {Finelli}, {Fiorini}, {Fynbo}, {Gal-Yam}, {Gendre}, {Ghisellini}, {Grado}, {Guidorzi}, {Hafizi}, {Hanlon}, {Hjorth}, {Izzo}, {Kiss}, {Kumar}, {Kuvvetli}, {Lavagna}, {Li}, {Longo}, {Lyutikov}, {Maio}, {Maiorano}, {Malcovati}, {Malesani}, {Margutti}, {Martin-Carrillo}, {Masetti}, {McBreen}, {Mignani}, {Morgante}, {Mundell}, {Nargaard-Nielsen}, {Nicastro}, {Palazzi}, {Paltani}, {Panessa}, {Pareschi}, {Pe'er}, {Penacchioni}, {Pian}, {Piedipalumbo}, {Piran}, {Rauw}, {Razzano}, {Read}, {Rezzolla}, {Romano}, {Ruffini}, {Savaglio}, {Sguera}, {Schady}, {Skidmore}, {Song}, {Stanway}, {Starling}, {Topinka}, {Troja}, {van Putten}, {Vanzella}, {Vercellone}, {Wilson-Hodge}, {Yonetoku}, {Zampa}, {Zampa}, {Zhang}, {Zhang}, {Zhang}, {Zhang}, {Antonelli}, {Bianco}, {Boci}, {Boer}, {Botticella}, {Boulade}, {Butler}, {Campana}, {Capitanio}, {Celotti}, {Chen}, {Colpi}, {Comastri}, {Cuby}, {Dadina}, {De Luca}, {Dong}, {Ettori}, {Gandhi}, {Geza}, {Greiner}, {Guiriec}, {Harms}, {Hernanz}, {Hornstrup}, {Hutchinson}, {Israel},
  {Jonker}, {Kaneko}, {Kawai}, {Wiersema}, {Korpela}, {Lebrun}, {Lu}, {MacFadyen}, {Malaguti}, {Maraschi}, {Melandri}, {Modjaz}, {Morris}, {Omodei}, {Paizis}, {P{\'a}ta}, {Petrosian}, {Rachevski}, {Rhoads}, {Ryde}, {Sabau-Graziati}, {Shigehiro}, {Sims}, {Soomin}, {Sz{\'e}csi}, {Urata}, {Uslenghi}, {Valenziano}, {Vianello}, {Vojtech}, {Watson}, and {Zicha}]{Theseus}
{Amati}, L.; {O'Brien}, P.; {G{\"o}tz}, D.; {Bozzo}, E.; {Tenzer}, C.; {Frontera}, F.; {Ghirlanda}, G.; {Labanti}, C.; {Osborne}, J.P.; {Stratta}, G.;  et~al.
\newblock {The THESEUS space mission concept: science case, design and expected performances}.
\newblock {\em Advances in Space Research} {\bf 2018}, {\em 62},~191--244,  \href{http://arxiv.org/abs/1710.04638}{{\normalfont [arXiv:astro-ph.IM/1710.04638]}}.
\newblock {\url{https://doi.org/10.1016/j.asr.2018.03.010}}.

\end{thebibliography}

%



\PublishersNote{}
\end{adjustwidth}
\end{document}